\documentclass[a4paper,12pt]{article}
\usepackage{amsmath}
\usepackage{amsthm}
\usepackage{amsfonts}
\usepackage{mathrsfs}
\usepackage{graphicx}
\usepackage{hyperref}
\usepackage{float}
\usepackage{braket}
\usepackage{float}
\usepackage{subcaption}
\usepackage{epsfig}
\usepackage{multirow}
\usepackage{xcolor}
\usepackage[sorting=none,style=numeric,giveninits=true]{biblatex}
\AtBeginBibliography{
  \DeclareNameAlias{author}{family-given}
}

\numberwithin{equation}{section}

\hypersetup{ pdftitle={Report 2},
pdfauthor={Ramy Fitrah Izzah}, 
pdfkeywords={nonlinear electrodynamics, cosmology, acceleration of the universe}, 
bookmarksnumbered, pdfstartview={FitH}, urlcolor=blue, }

\begin{document}
\pagestyle{plain}




\title{\LARGE\textbf{Inflation and Acceleration of the Universe from Nonminimal Coupling Gravity with Nonlinear Electrodynamics}}

\author{{Chilwatun Nasiroh, Ramy F. Izzah, Fiki T. Akbar, Bobby E. Gunara}\\ \\
$^{ }$\textit{\small
Theoretical Physics Laboratory, Theoretical High Energy Physics Research Division,}\\
\textit{\small Faculty of Mathematics and Natural Sciences,}\\
\textit{\small Institut Teknologi Bandung}\\
\textit{\small Jl. Ganesha no. 10 Bandung, Indonesia, 40132}\\
\small email: chilwatunnasiroh25@gmail.com, izzahramy@gmail.com, ftakbar@itb.ac.id, bobby@itb.ac.id}

\date{}

\maketitle




\begin{abstract} 
In this paper, we consider a nonminimal coupling model between gravity and nonlinear electrodynamics with cosmological constant. This cosmological model is designed to account for both the inflationary epoch of the early universe and the current phase of accelerated cosmic expansion. The nonlinear electrodynamic fields provide a mechanism for a graceful exit from the inflationary period, preventing the universe from entering an eternal inflation state. The addition of nonminimal coupling plays a crucial role in determining whether the transition from inflation to the subsequent cosmic phases results in accelerated or decelerated expansion. We compare the theoretical predictions of our model with recent observational data and other leading cosmological models, showing that our approach provides a viable and competitive explanation for key aspects of the universe's evolution. Our results suggest that this model offers a consistent and compelling framework to explain both early-time inflation and the late-time accelerated expansion of the universe, in line with current observations.
\end{abstract}




\section{Introduction}
Einstein's general theory of relativity is accepted as the best theory of gravity to explain our universe today. The application of this theory in the field of cosmology provides an overview of how the universe evolved. But there are several fundamental problems in standard cosmology, such as: why is our universe so flat (the flatness problem) and how could two different regions of spacetime be causally connected after the Big Bang (horizon problem). To overcome this problem, researchers proposed an inflation mechanism at the beginning of the universe. The theory of cosmic inflation was first proposed by Alan Guth in 1981. This theory proposes that in the early stages of its evolution, the universe underwent a brief period of exponential expansion, establishing initial conditions that could address several key issues in standard cosmology. Not only addressing the flatness and horizon problems \cite{guth1981inflationary,linde1982new}, the inflation theory also explains how primordial density fluctuations were generated, which later became the seeds for the large-scale structures we see in the universe today\cite{mukhanov1992theory,mukhanov2005physical,baumann2009cosmological}.

There are many inflation models that have been proposed, the most famous inflation model is inflation with a scalar field or slow-roll inflation, the observation results of PLANCK (2018) provide constraints and discuss the implications of slow-roll parameters in the context of inflation with a scalar field \cite{akrami2020planck}. The search for an inflation model that best fits the evolution of the universe is still an active area of research today. In addition, standard cosmology with Maxwell's electrodynamic sources based on Friedmann-Robertson-Walker (FRW) geometry produces singularities at certain times in the past.

One way to explain the universe experiencing inflation by avoiding singularities is to use nonlinear electrodynamics without modification of general relativity. The theory of nonlinear electrodynamics was first proposed by Max Born and Leopold Infeld in 1934 with the aim of eliminating point charge singularities in classical electrodynamics \cite{born1934foundations}. This absence of singularities is an interesting feature of nonlinear electrodynamics. Not only that, the Born-Infeld lagrangian will return to the Maxwell lagrangian for the case of low electromagnetic fields. Due to its unique features, nonlinear electrodynamics has many applications in the fields of cosmology \cite{de2002nonlinear,novello2007cosmological,vollick2008homogeneous,singh2018accelerating,sarkar2023emergent}, astrophysics \cite{harding2006physics,turolla2015magnetars}, and black holes \cite{moreno2003stability,breton2015thermodynamical,falciano2021entropy}.

Recent research shows that nonlinear electrodynamics (NLED) coupled with gravity can explain the inflation of the universe \cite{kruglov2017inflation,kruglov2020universe,ovgun2017inflation,benaoum2023inflation}. These studies use the Einstein-Hilbert lagrangian which is minimally coupled to nonlinear electrodynamics. There are many theories of gravity that are minimally coupled to other fields. Apart from minimal coupling, there are theories that couple the gravitational field with other fields using cross terms containing the curvature tensor. This theory is said to be nonminimally coupled. There are many fields that can be non-minimally coupled to gravity, for example the electromagnetic field \cite{balakin2005non}. In this paper we will use non-minimal coupling between gravity and nonlinear electrodynamics to obtain a solution for an inflationary universe.

Not only explaining inflation at early-time, our work also aim to explain the acceleration expansion at late-time. Many astronomical observations show that the universe is currently expanding at an increasingly rapid rate. These results are based on observations of Type Ia Supernovae (SN Ia) \cite{riess1998observational,perlmutter1999measurements}, CMB radiation (Cosmic Microwave Background) \cite{spergel2003first}, and large-scale structure \cite{daniel2008large}. In our study we add the cosmological constant ($\Lambda$) to the action of gravity with a nonminimal coupling model to obtain a solution for an accelerating universe. The cosmological model with the addition of the cosmological constant is known as the $\Lambda$CDM model. Since the $\Lambda$CDM model cannot produce two periods of accelerated expansion, one at early-time and another at late-time, we consider a gravity model that is nonminimally coupled with nonlinear electrodynamics to produce two periods of acceleration.

We perform a phase-space analysis by examining the Friedmann equations, which are analogous to the Hamiltonian of a one-dimensional particle \cite{ovgun2018falsifying,aguirregabiria2004note,lazkoz2005rigidity,szydlowski2007cosmological,mccrea1934newtonian,milne,mccrea1955significance,lima1998particle}. This analogy enables the system's dynamics to be described by an explicitly determined effective potential. This approach offers the advantage of utilizing potential functions instead of equations of state to study the universe's evolution \cite{szydlowski2007cosmological}. This framework provides extensive insights, including identifying critical points and assessing their stability \cite{wainwright1997dynamical}, which are crucial for understanding the universe's evolution over time.

Recent advancements have led to a significant increase in observational data, which provides valuable new insights into the expansion of the universe. The Cosmic Chronometer (CC) dataset has revealed the detailed structure of cosmic expansion. By measuring the ages of the largest galaxies, scientists have been able to directly determine the Hubble parameter \(H(z)\) at various redshifts \(z\) and establish a new standard cosmological probe. In this research, we utilize 44 CC measurements obtained through the differential age method \cite{bhardwaj2022corrected}, spanning a redshift range from \(0.07\) to \(2.36\) to place constraints on model parameters of constant.

The paper is organized as follows. In section \ref{sec:GR} we introduce the Lagrangian of nonlinear electrodynamics nonminimally coupled to the Ricci scalar. In section \ref{sec:hamiltonian} we perform the Hamiltonian formulation and examine the evolution of the universe. In section \ref{sec:dynamical system} we perform a detailed phase-space analysis. In section \ref{sec:observational constraints} we put constraints on the model parameters using observational Hubble data from Cosmic Chronometers. Finally, in section \ref{sec:conclusion}, we present our conclusions.


\section{Gravity and NLED with Nonminimal Coupling}
\label{sec:GR}

In this section, we shortly discuss a gravity and nonlinear electrodynamics with nonminimal coupling whose action is given by
\begin{equation}
S = \int \sqrt{-g} \left\lbrace \frac{1}{2\kappa^2} \left( R - 2\Lambda \right) + (1-\xi R) \mathcal{G}(\mathcal{F})\right\rbrace \text{d}^4x\;, \label{action}
\end{equation}
where $\kappa^{-1} = M_{\text{P}}$, $M_{\text{P}}, R, \xi, \Lambda$ denote the reduced Planck mass, the Ricci scalar, the nonminimal coupling constant and the cosmological constant, respectively. In this paper, the Greek index $\mu,\nu = 0,\ldots 3$ is a spacetime index, while the Latin index, $i,j=1,2,3$ is a spatial index. Moreover, we particularly take the form of nonlinear electrodynamics model which has been studied by \cite{kruglov2017inflation}:
\begin{equation}
\mathcal{G}(\mathcal{F}) = -\frac{\mathcal{F}}{(1+\beta\mathcal{F})^2} ~ , \label{nled}
\end{equation}
where $\beta$ is a positive real parameter, $\mathcal{F} \equiv \frac{1}{4} F_{\mu\nu} F^{\mu\nu}$, and $F_{\mu\nu} \equiv \partial_\mu A_\nu - \partial_\nu A_\mu$ is the field strength of the Abelian gauge field $A_\mu$. When $\beta \rightarrow 0$, Lagrangian \eqref{nled} reduces to the classical Maxwell’s electrodynamics $\mathcal{G}(\mathcal{F}) = -\mathcal{F}$. Nonlinear electrodynamics with a parameter $\beta$ studied by \cite{kruglov2017inflation} leads to universe inflation in the early universe where electromagnetic fields are the source of the gravitational field. After cosmic inflation, the universe transitions into a deceleration phase approaching Minkowski spacetime. This nonlinear electrodynamics model avoids the graceful exit problem that results from models with scalar fields, even though it is unable to explain the universe's late-time acceleration \cite{kruglov2017inflation}. 

The equations of motions  of gravity-NLED system are respectively given by
\begin{equation}
G_{\mu\nu} + g_{\mu\nu}\Lambda = \kappa^2 T_{\mu\nu} \;,
\end{equation}
and 
\begin{equation}
\partial_\mu \left(  \frac{\sqrt{-g}(1-\xi R)(1-\beta\mathcal{F})}{(1+\beta\mathcal{F})^3} F^{\mu\nu} \right) = 0 \;,
\end{equation}
where $G_{\mu\nu}$ is the Einstein tensor and $T_{\mu\nu}$ is the energy-momentum tensor of nonlinear electrodynamics field nonminimally coupled to gravity. We introduced the decomposition, 
\begin{equation}
T_{\mu\nu} =  T_{\mu\nu}^{M} + \xi T_{\mu\nu}^{(\xi)} \label{temnm}
\end{equation}
where $T_{\mu\nu}^{M}$ is the energy-momentum tensor of the matter (NLED) term and $T_{\mu\nu}^{\xi}$ is the energy-momentum tensor of due to the nonminimal coupling which can be expressed as follows 
\begin{eqnarray}
    T_{\mu\nu}^{(M)} &=& \frac{(1-\beta \mathcal{F}) }{(1+\beta \mathcal{F})^3}  F_{\mu\rho} F_\nu^\rho - g_{\mu\nu} \mathcal{G}(\mathcal{F}), \label{temm}\\
    T_{\mu\nu}^{(\xi)} &=& -\frac{(1-\beta \mathcal{F})R }{(1+\beta \mathcal{F})^3}  F_{\mu\rho} F_\nu^\rho - 2\left( G_{\mu\nu} - \nabla_\mu \nabla_\nu + g_{\mu\nu} \nabla^2\right) \mathcal{G}(\mathcal{F}), \label{temxi}
\end{eqnarray}
where $\nabla_\mu$ is a covariant derivative with respect to the metric $g_{\mu\nu}$ and $\nabla^2 \equiv \nabla_\mu \nabla^\mu$. It is important to note that energy-momentum tensor of the NLED field has non-zero trace in which is different from the standard Maxwell theory,
\begin{equation}
\mathcal{T}^{(M)} = \frac{8\beta \mathcal{F}^2}{(1+\beta \mathcal{F})^3}, \label{trm}
\end{equation}
and the trace of the energy momentum tensor due to nonminimal coupling is
\begin{equation}
\mathcal{T}^{(\xi)} = -\frac{4\mathcal{F}R(1-\beta \mathcal{F})}{(1+\beta \mathcal{F})^3} + 2\left(R - 3g^{\mu\nu} \nabla_\mu \nabla_\nu \right) \mathcal{G}(\mathcal{F}). \label{trxi}
\end{equation}

In particular, throughout this paper, we will consider the gravity-NLED system on a homogeneous and isotropic spatially flat universe described by the Friedmann-Robertson-Walker (FRW) metric
\begin{equation}
\text{d}s^2 = -\text{d}t^2 +a(t)^2 (\text{d}x^2 + \text{d}y^2 + \text{d}z^2), \label{frw}
\end{equation}
where $a(t)$ is the scale factor as a function of the cosmic time.

Then, we add an assumption that there are dominant stochastic magnetic fields in the cosmic background, and their wavelengths are much smaller than the curvature. Consequently, the averaging of the electromagnetic field can be used as a source for the Einstein equation (see Ref. \cite{tolman1930temperature}). Thus, the spatially averaged electromagnetic fields must satisfy the following conditions:
\begin{equation}
    \begin{aligned}
        & \braket{E} = \braket{B} = 0, \qquad\braket{E_i B_j} = 0, \\
        & \braket{E_i E_j} = 0, \qquad \braket{B_i B_j} = \frac{1}{3} B^2 g_{ij}. \label{magnetic}
    \end{aligned}
\end{equation}
Note that in the rest of the paper, the averaging brackets $\braket{ \cdot }$ are omitted  for simplicity. 

Therefore, the energy-momentum tensor in Eq. \eqref{temnm} can be written in the perfect fluid form \cite{novello2007cosmological},
\begin{equation}
    T_{\mu\nu} = (\rho + p)U_{\mu} U_{\nu} + g_{\mu\nu} p,
\end{equation}
where $U^{\mu}$ is the 4-velocity of the fluid rest frame, $\rho$ and $p$ are the energy density and the pressure, respectively which given by  
\begin{eqnarray}
    \rho &=& \frac{2B^2}{(2+\beta B^2)^2} - \frac{12\xi B^2}{(2+\beta B^2)^2} \left( \frac{\dot{a}^2}{a^2}\right) - \frac{24\xi B\dot{B}(2-\beta B^2)}{(2+\beta B^2)^3} \left(\frac{\dot{a}}{a}\right), \label{rho} \\
    p &=& \frac{2B^2 (2-7\beta B^2)}{3(2+\beta B^2)^3} - \frac{8\xi B^2(2-3\beta B^2)}{(2+\beta B^2)^3} \left( \frac{\ddot{a}}{a}\right) - \frac{4\xi B^2(6-5\beta B^2)}{(2+\beta B^2)^3} \left(\frac{\dot{a}^2}{a^2}\right)  \nonumber\\
    & & + \frac{16\xi B\dot{B}(2-\beta B^2)}{(2+\beta B^2)^3} \left( \frac{\dot{a}}{a}\right) + 8\xi \left( \frac{B\ddot{B} (4-\beta^2 B^4) + \dot{B}^2(3\beta^2 B^4 - 16\beta B^2 + 4)}{(2+\beta B^2)^4}\right)\;, \nonumber\\
    \label{p}
\end{eqnarray}
where dot denotes the derivative with respect to the cosmic time, $t$.

The conservation of the energy-momentum tensor, $\nabla^\mu T_{\mu\nu} = 0$, gives the relation
\begin{equation}
\dot{\rho} + 3 \frac{\dot{a}}{a} (\rho + p) = 0 ~ . \label{f4}
\end{equation}
Substituting equation \eqref{rho} and equation \eqref{p} into equation \eqref{f4} yields 
\begin{equation}
\left( \dot{B} + 2 B \frac{\dot{a}}{a}\right) \left[ 1- 6\xi  \left( \frac{\ddot{a}}{a} + \frac{\dot{a}^2}{a^2}\right)\right] = 0 \;
\end{equation}
which implies,
\begin{equation}
B(t) = \frac{B_0}{a^2(t)} \;, \label{B}
\end{equation}
where $B_0 = B(t_0)$ is the value of magnetic field when the scale factor $a(t_0) =1$. This result shows that the magnetic field decreases as the scale factor $a(t)$ increase implying that the universe was initially filled with a very high stochastic magnetic field during the early stage of the evolution of the universe that driving inflation to occur. The relationship between the decay of the magnetic field and the expansion of the universe has significant implications for our understanding of cosmic inflation. High initial magnetic fields can influence the dynamics of inflation and these fields can generate primordial density perturbations, which are seeds for the large-scale structure of the universe we observed today \cite{turner1988inflation}. Moreover, the decay of magnetic fields in an expanding universe provides insights into magnetogenesis \cite{kahniashvili2016evolution}, the process by which cosmic magnetic fields are generated, which remains an active area of research in cosmology.

In the model, we have the first and second Friedmann equations are given by
\begin{eqnarray}
\left( \frac{\dot{a}^2}{a^2}\right) &=& \frac{\kappa^2\rho_{\text{total}}}{3} + \frac{\Lambda}{3} ~ , \label{f1}\\ 
3\frac{\ddot{a}}{a} &=& -\frac{\kappa^2}{2} (\rho_\text{total} + 3 p_{\text{total}}) + \frac{\Lambda}{3}\:,  \label{f2}
\end{eqnarray}
respectively, where $\rho_{\text{total}} = \rho_{\text{eff}} + \rho_{\Lambda}$ denotes the total energy density with, 
\begin{eqnarray}
\rho_{\text{eff}}  & = & \frac{2a^4 B^2_0 (2a^4 +\beta B^2_0)}{(2a^4+\beta B^2_0)^3 - 4\xi\kappa^2 a^4 B^2_0 (6a^4 -5\beta B^2_0)}, \label{rhoeffL} \\
\rho_{\Lambda} & = &  \frac{ (2a^4 + \beta B^2_0)^3}{(2a^4+\beta B^2_0)^3 - 4\xi\kappa^2 a^4 B^2_0(6a^4-5\beta B^2_0)}\frac{\Lambda}{\kappa^2}\;, \label{rhoLam}
\end{eqnarray}
and $p_{\text{total}} = p_{\text{eff}} + p_{\Lambda}$ denotes the total pressure with
\begin{eqnarray}
p_{\text{eff}} & = & \frac{2a^4 B^2_0 }{3} \left[ \frac{(2a^4 - 7\beta B^2_0)(2a^4 + \beta B^2_0)^3 - 4\xi \kappa^2 a^4 B^2_0 (5\beta^2 B^4_0 + 68a^4 \beta B^2_0 - 12a^8)}{\left[ (2a^4+\beta B^2_0)^3 - 4\xi\kappa^2 a^4 B^2_0 (6a^4 - 5\beta B^2_0)\right]^2}\right], \nonumber \\
\label{peffL} \\
p_\Lambda  & = &  - \frac{\Lambda}{3\kappa^2}(2a^4 +\beta B^2_0)^2 \left[\frac{3(2a^4 + \beta B^2_0)^4 - 4\xi\kappa^2 a^4 B^2_0 (5\beta^2 B^4_0 - 140 a^4 \beta B^2_0 + 84a^8) }{\left[ (2a^4+\beta B^2_0)^3 - 4\xi\kappa^2 a^4 B^2_0 (6a^4 - 5\beta B^2_0)\right]^2}\right]. \nonumber \\
\label{pLam}
\end{eqnarray}
Taking the limit for early evolution, $a(t) \rightarrow 0$, and the limit for late evolution, $a(t) \rightarrow \infty$, we obtain 
\begin{eqnarray}
\lim_{a(t) \rightarrow 0} \rho_{\text{eff}} (t) = \lim_{a(t) \rightarrow 0}  p_{\text{eff}} = 0, &\quad& \lim_{a(t) \rightarrow 0} \rho_\Lambda = \frac{\Lambda}{\kappa^2}, \quad \lim_{a(t) \rightarrow 0} p_\Lambda = -\frac{\Lambda}{\kappa^2} ~ , \\
\lim_{a(t) \rightarrow \infty} \rho_{\text{eff}} (t) = \lim_{a(t) \rightarrow \infty}  p_{\text{eff}} = 0, &\quad& \lim_{a(t) \rightarrow \infty} \rho_\Lambda = \frac{\Lambda}{\kappa^2}, \quad \lim_{a(t) \rightarrow \infty} p_\Lambda = -\frac{\Lambda}{\kappa^2} ~ .
\end{eqnarray}
These results show  that there are no singularities in energy density and pressure in the limits $a(t) \rightarrow 0$ and $a(t) \rightarrow \infty$. This absence of singularities is a one of the important feature of cosmological models with nonlinear electrodynamics. This characteristic has also been demonstrated in the works of \cite{kruglov2017inflation, kruglov2020universe, ovgun2017inflation} and \cite{benaoum2023inflation} using different models of nonlinear electrodynamics.

\begin{figure*}[ht!]
	\centering
\includegraphics[width=.7\linewidth]{"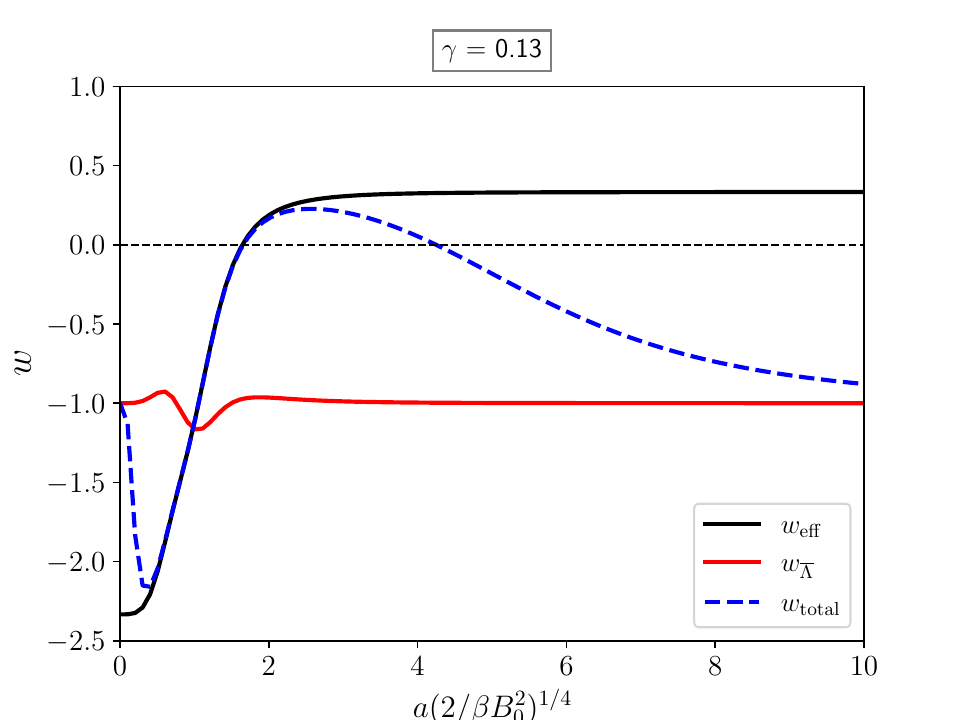"}
	\caption{The evolution of the equation of state parameters $w_{\text{eff}}$, $w_{\overline{\Lambda}}$, and $w_{\text{total}}$ versus scale factor $x = a \left( 2/\beta B_0^2 \right)^{\frac{1}{4}}$ for $\gamma=0.13$. The value of $w_{\text{total}}$ (dashed blue line) and $w_{\overline{\Lambda}}$ (solid red line) approach -1 at $x\rightarrow 0$ and $x \rightarrow \infty$. It is shown that the nonminimal coupling model with a cosmological constant plays a crucial role in the universe's early and late-time evolution, compared to the model with nonminimal coupling and nonlinear electrodynamics alone (solid black line). It also influenced the evolution of $w_{\overline{\Lambda}}$ (solid red line) as indicated by the fluctuations.
 }
	\label{wfigL}
\end{figure*}

Now, we could consider a case where the universe as consisting of two-component matters with the equation of state:
\begin{equation}
    p_{\text{total}} = p_{\text{eff}} + p_{\Lambda} = w_{\text{eff}}\rho_{\text{eff}} + w_{\Lambda}\rho_{\Lambda} = w_{\text{total}}\rho_{\text{total}}\;,
\end{equation}
where $w_{\text{eff}}$ and $w_{\Lambda}$ respectively denote the equation of state parameters for NLED with nonminmal coupling and cosmological constant which given by
\begin{eqnarray}
w_{\text{eff}} &=& \frac{1}{3(2a^4+\beta B_0^2)} \left[ \frac{(2a^4-7\beta B_0^2)(2a^4+\beta B_0^2)^3 - 4\xi \kappa^2 a^4B_0^2 (5\beta^2 B^4_0 + 68a^4\beta B^2_0 - 12a^8)}{(2a^4 + \beta B^2_0)^3 - 4\xi\kappa^2 a^4 B^2_0(6a^4-5\beta B^2_0)}\right], \nonumber\\
\label{weffa} \\
w_{\Lambda} &=& - \frac{1}{3(2a^4+\beta B^2_0)}\left[ \frac{ 3(2a^4+\beta B^2_0)^4 - 4\xi\kappa^2 a^4 B^2_0 (5\beta^2 B^4_0 - 140 a^4 \beta B^2_0 + 84a^8) }{ (2a^4+\beta B^2_0)^3 - 4\xi\kappa^2 a^4 B^2_0(6a^4-5\beta B^2_0)}\right]. \label{wLa}  
\end{eqnarray}
Similar as above, taking  the limit for early evolution, $a(t) \rightarrow 0$, and the limit for late evolution, $a(t) \rightarrow \infty$, we get
\begin{eqnarray}
\lim_{a(t) \rightarrow 0} w_{\text{eff}} &=& -\frac{7}{3}, \qquad \lim_{a(t) \rightarrow 0} w_{\Lambda} = -1, \qquad \lim_{a(t) \rightarrow 0} w_{\text{total}} = -1\\ 
\lim_{a(t) \rightarrow \infty} w_{\text{eff}} &=& \frac{1}{3}, \qquad \lim_{a(t) \rightarrow \infty} w_{\Lambda} = -1 \qquad \lim_{a(t) \rightarrow \infty} w_{\text{total}} = -1. \label{wlate}
\end{eqnarray}
Figure \ref{wfigL} shows the evolution of $w_{\text{eff}}$ and $w_{\Lambda}$ with respect to the scale factor $x = a \left( 2/\beta B_0^2 \right)^{\frac{1}{4}}$. The accelerated expansion of the universe occurs if the parameters of the equation of state are less than $-1/3$. From equation \eqref{wlate}, we can see that $w_{\text{total}}$ at late-time is $-1$, indicating that the cosmological constant dominates late-time evolution and can produce a late-time accelerating universe. The NLED with the nonminimal coupling term plays a crucial role in the early evolution of the universe and is related to the inflationary scenario of the universe


\section{The Evolution of the Universe}
\label{sec:hamiltonian}
In this section, we discuss about the solution of the Friedmann equation and examine how our model describes the evolution of the universe. The main goal of this study is to understand the behavior and dynamics of the cosmological model we are exploring. In particular, we use the well known result that the first Friedmann equation can be written as one dimensional Hamiltonian form with the scale factor will be considered as a dynamical variable in this system. This allows us to explore the system's dynamics and identify important critical points. By analyzing the stability of these points, we can understand how the scale factor evolves throughout different cosmological eras. 

Making use of Eqs. \eqref{rhoeffL}, \eqref{rhoLam} and \eqref{B} to the first Friedmann equation yields 
\begin{eqnarray}
\frac{\dot{a}^2}{a^2} &=& \frac{1}{3}\frac{2\kappa^2a^4 B_0^2 (2a^4+\beta B_0^2) + \Lambda(2a^4+\beta B_0^2)^3}{(2a^4+\beta B_0^2)^3 - 4\xi\kappa^2a^4 B_0^2(6a^4 -5\beta B_0^2)}. \label{adot}
\end{eqnarray}
We define a dimensionless parameters, $b\equiv(\sqrt{\beta}B_0)/\sqrt{2}$, $\gamma\equiv \xi \kappa^2/\beta$, $\overline{\Lambda}\equiv\beta \Lambda/\kappa^2$.
We introduce a re-scaled time variable, $\tau = (\kappa t\left|B_0\right|)/\sqrt{6}$, and re-scaled scale factor, $x = a/\sqrt{b}$. Then we can write the Eq. \ref{adot} as
\begin{equation}
\frac{dx}{d\tau} = \frac{x}{b} \sqrt{ \frac{x^4 (x^4+1) + \bar{\Lambda}(x^4+1)^ 3}{(x^4+1)^3 - 2 \gamma x^4 (3x^4 -5)}}, \label{num}
\end{equation}
The Eq. \eqref{num} cannot be solved exactly, therefore we will take the approximation and numerical solution of Eq. \eqref{num}. The approximation to the exact solution can be calculated using the Taylor expansion series around $\overline{\Lambda}\xrightarrow[]{}0$ and $\gamma\xrightarrow[]{}0$. Hence, we obtain
\begin{equation}
\frac{1}{b}d\tau \approx \left[\left(\frac{1}{x^3}+x \right) + \frac{5x-3x^5}{(1+x^4)^2}\gamma-\frac{(1+x^4)^3}{2x^7}\overline{\Lambda} \right] dx \label{aproksimasi}
\end{equation}
Integrating the equation with the initial value $x(0)=1$, we obtain,
\begin{equation}
\tau = b \left[ \frac{x^4-1}{2x^2} + \left( \frac{2x^2}{x^4+1}+\frac{1}{2}\arctan{x^2} -1 - \frac{\pi}{8} \right)\gamma+\left(\frac{1+9x^4-9x^8-x^{12}}{12x^6}\right)\overline{\Lambda} \right]\:. \label{t}
\end{equation}
Then the real approximation solution to Eq. \eqref{num} is given by
\begin{eqnarray}
x(\tau)&\approx&\frac{\sqrt{\tau + \sqrt{\tau^2+b^2}}}{\sqrt{b}} + \frac{\tau (\tau^2 + 3b^2)\sqrt{b \left(\tau + \sqrt{\tau^2 + b^2} \right)}}{3b^3\sqrt{\tau^2+b^2}} \overline{\Lambda} \nonumber\\
& &+ \frac{\sqrt{b\left(\tau + \sqrt{\tau^2+b^2}\right)}}{16\sqrt{\tau^2+b^2}} \left(8+\pi - \frac{8b}{\sqrt{\tau^2+b^2}}-4 \arctan{\left[ \frac{\tau +\sqrt{\tau^2+b^2}}{b}\right]} \right)\gamma \label{solusiaproksimasi} \nonumber\\
\end{eqnarray}

Equation \eqref{solusiaproksimasi} allows us to study the evolution of the universe. In the early evolution, $\tau\xrightarrow[]{}0$, a model of gravity coupled nonminimally with nonlinear electrodynamics and with the cosmological constant $\Lambda$ could explain the inflation of the universe. Meanwhile, at the late stage of the evolution, $\tau \xrightarrow{} \infty$, the solution \eqref{solusiaproksimasi} becomes $x\xrightarrow[]{}\sqrt{2\tau/b}\left(1+\overline{\Lambda}\tau^2/(3b^2) \right)$ corresponding to the late-time acceleration. This late time solution shows a polynomial growth of the scale factor over time, reflecting the influence of cosmological constant $\Lambda$ as the dominant form of energy in this epoch. Thus, the nonminimal coupling model with the cosmological constant can provide a solution for a universe that experiences inflation in the early universe and acceleration in the late-time. 

\begin{figure}[ht]
	\centering
	\begin{subfigure}{0.48\textwidth}
		\includegraphics[width=\textwidth]{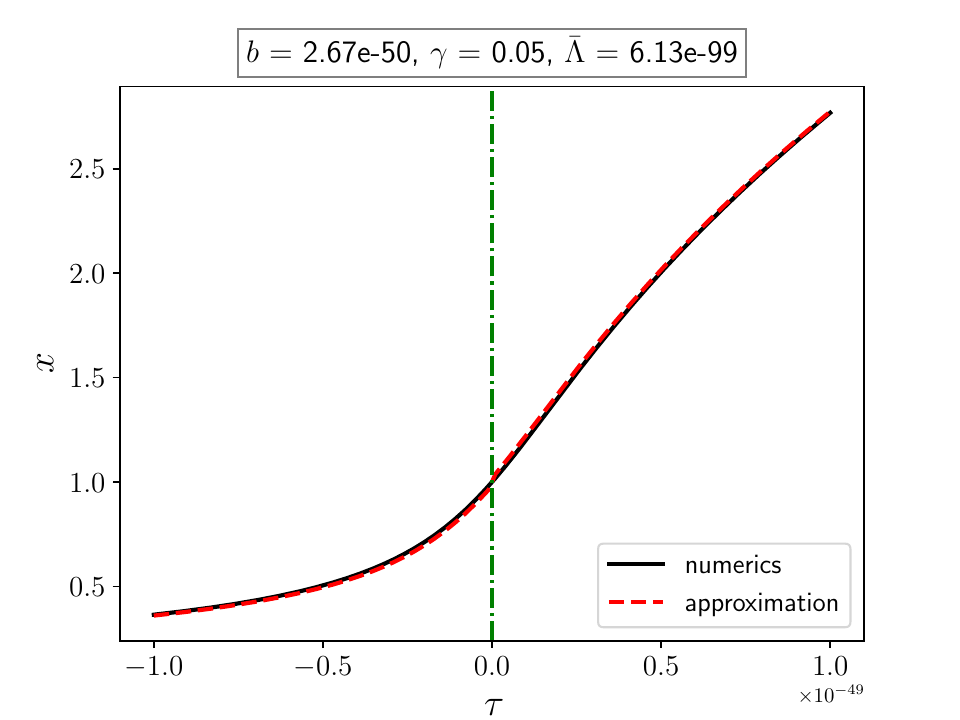}
		\caption{}
	\end{subfigure}
	\begin{subfigure}{0.48\textwidth}
		\includegraphics[width=\textwidth]{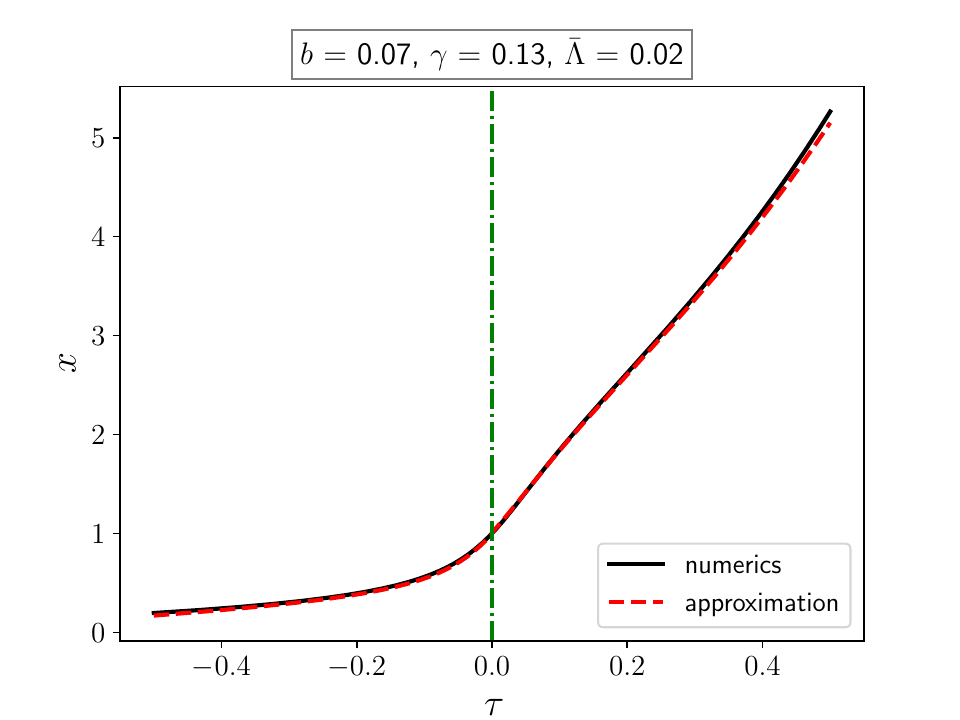}
		\caption{}
	\end{subfigure}
	\caption{Plot $x$ vs. $\tau$ as a approximation and numerical solution of Eq. \eqref{num} for the choice of values (a) $\gamma=0.05, b=2.67 \times 10^{-50}$, $\overline{\Lambda}=6.13 \times 10^{-99}$, and (b) $\gamma=0.13$, $b=0.07$, $\overline{\Lambda}=0.02$. }
 \label{afig}
\end{figure}

We also solve Eq. \eqref{num} numerically and the plot the comparison with the approximation solution \eqref{solusiaproksimasi} in Fig. \ref{afig}. From the figure, it can be seen that the approximate solution matches the numerical solution for $\gamma \xrightarrow[]{}0$ and $\overline{\Lambda}\xrightarrow[]{}0$.

We can express the first Friedmann equation \eqref{f1} in the form of a Hamiltonian as:
\begin{equation}
\mathcal{H} = \frac{1}{2}\dot{a}^2 + U_{\text{eff}}(a) = 0\:, \label{Hamiltonian}
\end{equation}
where the effective potential $U_{\text{eff}}(a)$ is given by:
\begin{eqnarray}
U_{\text{eff}}(a) = -\frac{1}{6}\left[ \frac{2\kappa^2a^4 B_0^2 (2a^4+\beta B_0^2) + \Lambda(2a^4+\beta B_0^2)^3}{(2a^4+\beta B_0^2)^3 - 4\xi\kappa^2a^4 B_0^2(6a^4 - 5\beta B_0^2)} \right]. \label{Ueff}
\end{eqnarray}

In this formulation, the dynamics of the universe are analogous to the motion of a particle with unit mass in a one-dimensional potential field. The scale factor $a(t)$, which measures the expansion or contraction of the universe, corresponds to the particle's position. Just as the motion of a particle is influenced by the potential it experiences, the evolution of the universe is governed by the Hamiltonian, where the effective potential $U_{\text{eff}}(a)$ plays a key role. This analogy provides a useful framework for understanding cosmic evolution in terms of more familiar classical mechanics.

To explore the system further, we can reformulate the equation into two coupled first-order differential equations:
\begin{eqnarray}
\frac{dx}{d\tau} = y, \quad \frac{dy}{d\tau} = -\frac{\partial W(x)}{\partial x}, \label{dydt}
\end{eqnarray}
where $y = \dot{a}\sqrt{6}/(\kappa B_0 \sqrt{b})$ and the potential function $W(x)$ is:
\begin{equation}
W(x) = \frac{6U_{\text{eff}}(x)}{\kappa^2 B_0^2} = -\frac{1}{2b^2}\left[\frac{x^6(x^4+1) + \overline{\Lambda}x^2(x^4+1)^3}{(x^4+1)^3 - 2\gamma x^4(3x^4 - 5)} \right]. \label{Wx}
\end{equation}

The form \eqref{dydt} transforms the problem into a dynamical system, which allows us to apply dynamical systems methods to investigate all possible evolutionary scenarios, given any initial conditions. In fact, the behavior of the universe can be directly derived from the properties of the potential function $W(x)$. However, it is important to note that the effective potential is only valid for $\gamma \leq 2.15$; beyond this threshold, the potential becomes singular.

The shape of the effective potential, $W(x)$, plays a crucial role in determining the evolution of the universe. When the potential decreases with respect to the variable $x$, it indicates that the universe will undergo an accelerated expansion. Conversely, when the potential increases, it implies a decelerating phase. These phases correspond to the different epochs in the universe’s history, with deceleration occurring after inflation and acceleration emerging again at late times.

According to cosmological observations, the universe has been experiencing accelerated expansion only relatively recently, during the late-time epoch, following a long period of deceleration after the inflationary phase \cite{perlmutter1999measurements,riess1998observational}. Therefore, any comprehensive cosmological model must be capable of describing both phases—decelerating and accelerating—in order to provide a complete picture of cosmic evolution.

Since the second equation in Eq. (\ref{dydt}) describes the dynamics of the scale factor's evolution, it is directly linked to the universe's acceleration or deceleration. The critical points of this equation contain valuable information about the transitions between these two phases, helping us to identify when the universe shifts from a period of deceleration to one of acceleration. The interplay between these epochs is critical to understanding the overall dynamics of the universe, as outlined in various studies of dynamical systems in cosmology \cite{copeland2006dynamics}.

The non-trivial critical points satisfy the following polynomial equation:
\begin{eqnarray}
0 &=& -\overline{\Lambda} - x^4(3 + 6\overline{\Lambda} - 10\gamma \overline{\Lambda}) - x^8(8 + 15\overline{\Lambda} + 10\gamma + 48 \gamma \overline{\Lambda}) \nonumber\\
&& - 2x^{12}(3 + 10\overline{\Lambda} + 18\gamma + 54 \gamma \overline{\Lambda}) - x^{16}(15\overline{\Lambda} - 6\gamma + 32 \gamma \overline{\Lambda}) \nonumber\\
&& + x^{20}(1 - 6\overline{\Lambda} + 18 \gamma \overline{\Lambda}) - \overline{\Lambda}x^{24}. \label{Wx'}
\end{eqnarray}

Since the polynomial is basically a sextic equation and both $\overline{\Lambda}$ and $\gamma$ are assumed to be positive real numbers, the roots of the polynomial can either include two positive real roots or none. If no positive real roots exist, this suggests the absence of a deceleration epoch, meaning the universe will continue accelerating after the inflationary period, without any intermediate slowdown. However, if there are two positive real roots, the universe will experience a phase of deceleration following inflation, before eventually transitioning back to an accelerated expansion. 

\begin{figure}[ht]
  \centering
  \includegraphics[width=0.7\textwidth]{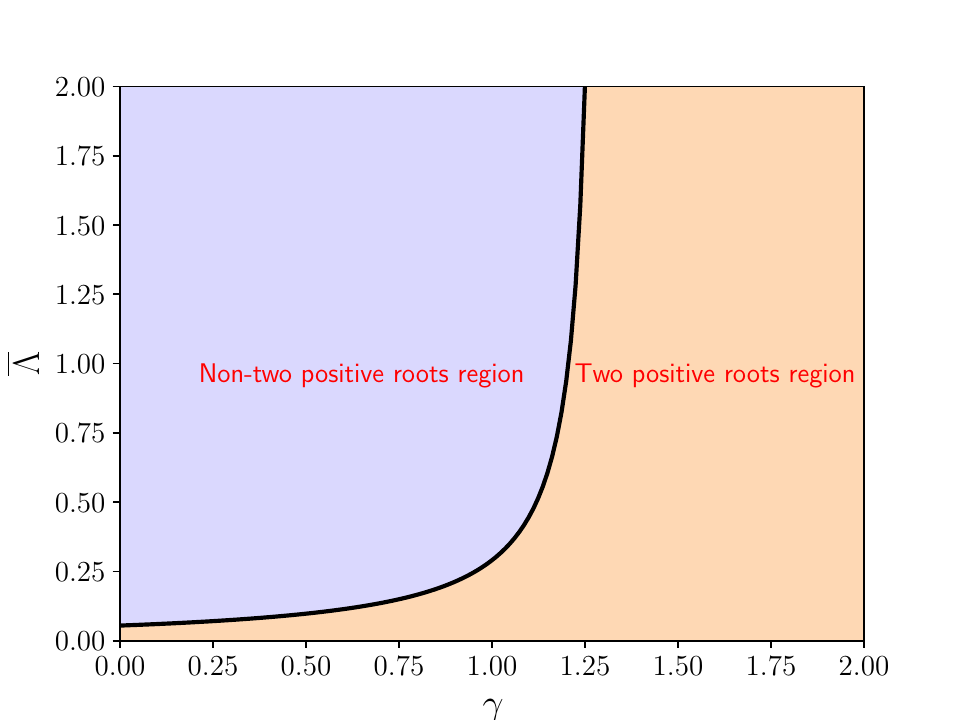}
  \caption{The plot represents the region in the $\overline{\Lambda}$ vs. $\gamma$ parameter space where two positive roots exist and where non-two positive roots are present. The boundary between these regions is of particular interest as it helps in understanding the behavior of the model under different conditions.}
  \label{fig:2roots}
\end{figure}

The parameter space for these two scenarios is illustrated in Fig. \ref{fig:2roots}. The boundary separating these regions is described by:
\begin{eqnarray}
\overline{\Lambda}_{\text{max}} \approx 0.0547 - 0.0353 \gamma + (0.0732 + 0.0279 \gamma) \tan(1.214 \gamma). \label{Lambdamax}
\end{eqnarray}
In this paper, we focus on the second model, where the universe undergoes a phase of deceleration after inflation, followed by a return to accelerated expansion. Thus, we assume that $\overline{\Lambda} \leq \overline{\Lambda}_{\text{max}}$. Under this assumption, the positive real roots are given by:
\begin{eqnarray}
x_1 &\approx& \frac{16}{(3)^{\frac{7}{4}}} \overline{\Lambda} + \frac{3^{\frac{1}{4}}}{4}(4 + \gamma), \label{426}\\
x_2 &\approx& -\frac{3}{2} \overline{\Lambda}^{\frac{3}{4}} + \left(\overline{\Lambda}\right)^{-\frac{1}{4}}. \label{428}
\end{eqnarray}
It is important to note that in the absence of a cosmological constant, the second root, $x_2$, tends toward infinity, leaving only one finite positive root. In this case, after the inflationary period, the universe would experience eternal deceleration, failing to generate the late-time acceleration as discussed in \cite{kruglov2017inflation}. This conclusion is further supported by Fig. \ref{wfigL}, where $w_\text{eff}$ exhibits positive values at late times, indicating an absence of late-time acceleration.

The inflation phase takes place within the range \(x(0) = 1\) to \(x_i = x_1\). Using Eq. \eqref{t}, we determine the duration of inflation, as follows:
\begin{eqnarray}
\tau_i &=& b \left[\frac{1}{\sqrt{3}} + \frac{44}{9\sqrt{3}} \overline{\Lambda} + \gamma \left(-1 + \frac{5}{2\sqrt{3}} + \frac{\pi}{24}\right)\right]. \label{Ti}
\end{eqnarray}
After the inflationary phase concludes, the universe begins to decelerate. Deceleration occurs within the range \(x_i < x < x_d\), where \(x_i = x_1\) and \(x_d = x_2\). Therefore, the duration of deceleration is given by 
\begin{equation}
\tau_d = b\left[-\frac{1}{\sqrt{3}} + \frac{5}{12\sqrt{\overline{\Lambda}}} - 2\sqrt{\overline{\Lambda}} - \frac{44}{9\sqrt{3}} \overline{\Lambda} + \frac{1}{12} \gamma \left(-10\sqrt{3} + \pi\right)\right]. \label{Td}
\end{equation}
The duration of the inflation and deceleration periods depend on the values of $b, \gamma$, and $\overline{\Lambda}$. Therefore, these duration can serve as constraints for determining the model parameters, which will be discussed in the following section.

\section{Dynamical System}
\label{sec:dynamical system}
\noindent In order to study the stability of the critical points, we consider linear perturbations around them. The nature of each critical point is determined by the eigenvalues of the linearization matrix $J$, which satisfy the equation $\lambda^2+\det{J}=0$, where $J = \frac{\partial^2}{\partial x^2} W(x)\big|_{x = x_0}$, with $W(x)$ evaluated at the critical point $x_0$.

\noindent Critical point 1:$\left(\left[3^{\frac{1}{4}}+\frac{16}{(3)^{\frac{7}{4}}}\overline{\Lambda}+\frac{3^{\frac{1}{4}}}{4}\gamma\right],0 \right)$ has the following Jacobian matrix:
\begin{eqnarray}
J &=&\begin{pmatrix}
                0 & 1\\
                \left(-\frac{9}{16b^2}+\frac{6\overline{\Lambda}}{b^2}-\frac{63}{64b^2}\gamma\right) & 0
    \end{pmatrix}. \label{J1}\
\end{eqnarray}
The eigenvalues of \eqref{J1} are
\begin{eqnarray}
\lambda_1 =  \left(\frac{3}{4}-4\overline{\Lambda}+\frac{21}{32}\gamma \right) \sqrt{-\frac{1}{b^2}}, \quad \lambda_2 =- \left(\frac{3}{4}-4\overline{\Lambda}+\frac{21}{32}\gamma \right) \sqrt{-\frac{1}{b^2}}. \label{eigenvalues cp1}
\end{eqnarray}
Thus, the critical point 1 is a centre. 

\begin{figure}[ht]
	\centering
	\begin{subfigure}{0.48\textwidth}
		\includegraphics[width=\textwidth]{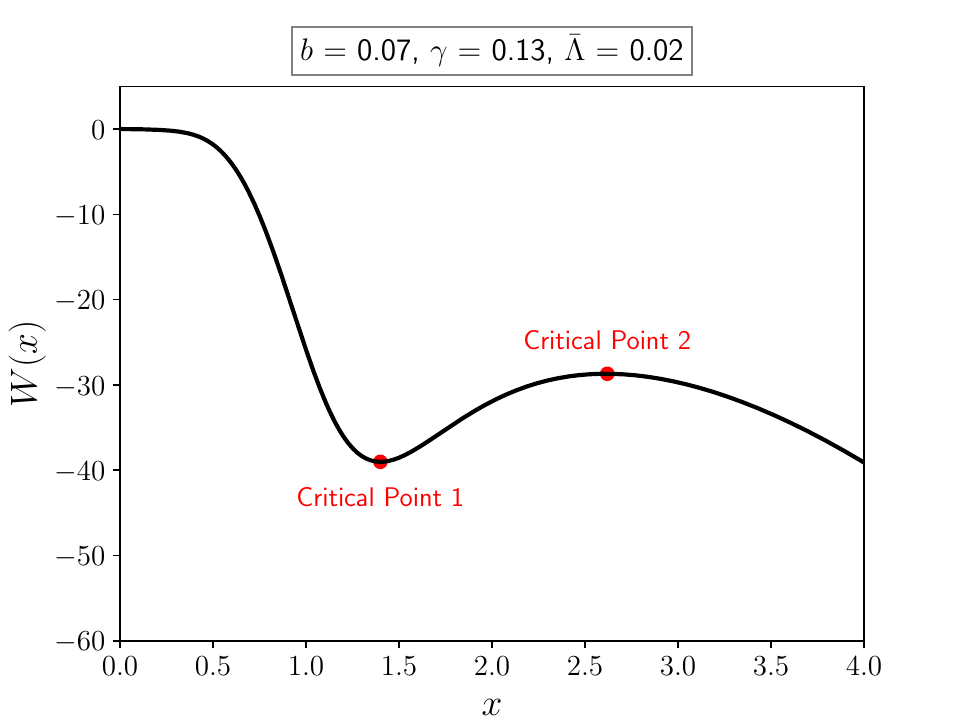}
		\caption{}
	\end{subfigure}
	\begin{subfigure}{0.48\textwidth}
		\includegraphics[width=\textwidth]{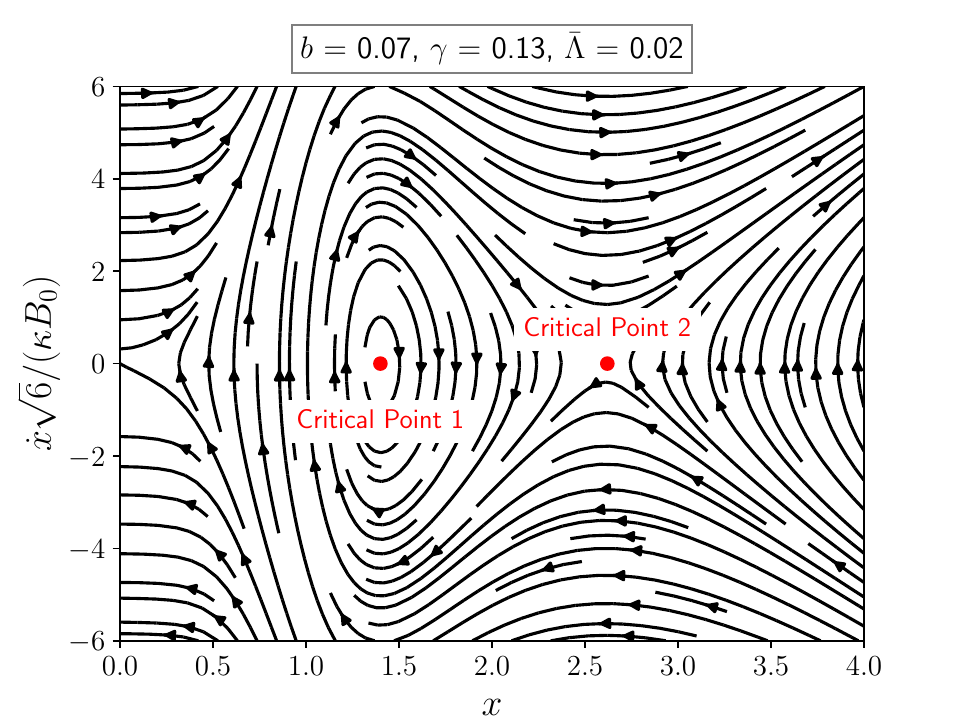}
		\caption{}
	\end{subfigure}
	\caption{The plot of (a) effective potential $W(x)$ and (b) the phase portrait correspond to critical point 1 and 2 for the choice $ b=0.07, \gamma=0.13$, and $\overline{\Lambda}=0.02$. }
 \label{fig:kasus3}
\end{figure}

\noindent Similarly, critical point 2:$\left(\left[-\frac{3}{2}\overline{\Lambda}^{\frac{3}{4}}+\left(\overline{\Lambda} \right)^{-\frac{1}{4}}\right],0 \right)$ has a Jacobian matrix:
\noindent \begin{eqnarray}
J &=&\begin{pmatrix}
                0 & 1\\
                \left[\frac{4\overline{\Lambda}}{b^2}-\frac{24\overline{\Lambda}^2}{b^2}\right] & 0
    \end{pmatrix} ,\label{J2}\
\end{eqnarray}
with eigenvalues
\begin{eqnarray}
\lambda_1=\left(2-6\overline{\Lambda} \right)\sqrt{\frac{\overline{\Lambda}}{b^2}}, \quad \lambda_2=-\left(2-6\overline{\Lambda}\right)\sqrt{\frac{\overline{\Lambda}}{b^2}}. \
\end{eqnarray}
Thus, the critical point 2 is a saddle. The plot of these critical points is represented in Fig. \ref{fig:kasus3}. It can be seen that the scale factor increases in the time from acceleration phase to deceleration phase before finally reaching the late-time acceleration phase.

\noindent Since we consider that the scale factor is non-negative ($x> 0$), we can define the new variables:
\begin{eqnarray}
x=e^{u}, \quad y=v,
\end{eqnarray}
which takes values on the real line. A time variable, denoted by $\eta \equiv \int \exp{(-u)} d\tau$, is introduced to represent the evolution of the system across all real numbers. Therefore, the dynamic equation yields:
\begin{eqnarray}
\frac{du}{d\eta}=v, \quad \frac{dv}{d\eta}=-\frac{\partial W(u)}{\partial u}, \label{47}\
\end{eqnarray}
where $W(u)=6U_{eff}(u)/(\kappa^2 B_0^2b)$. Therefore, we have
\begin{eqnarray}
W(u)&=&-\frac{1}{2b^2}\left[\frac{e^{6u}(e^{4u}+1)+\overline{\Lambda}e^{2u}(e^{4u}+1)^3}{(e^{4u}+1)^3-2\gamma  e^{4u}(3e^{4u}-5)} \right], 
\end{eqnarray}

\begin{figure}[ht]
  \centering
  \includegraphics[width=0.6\textwidth]{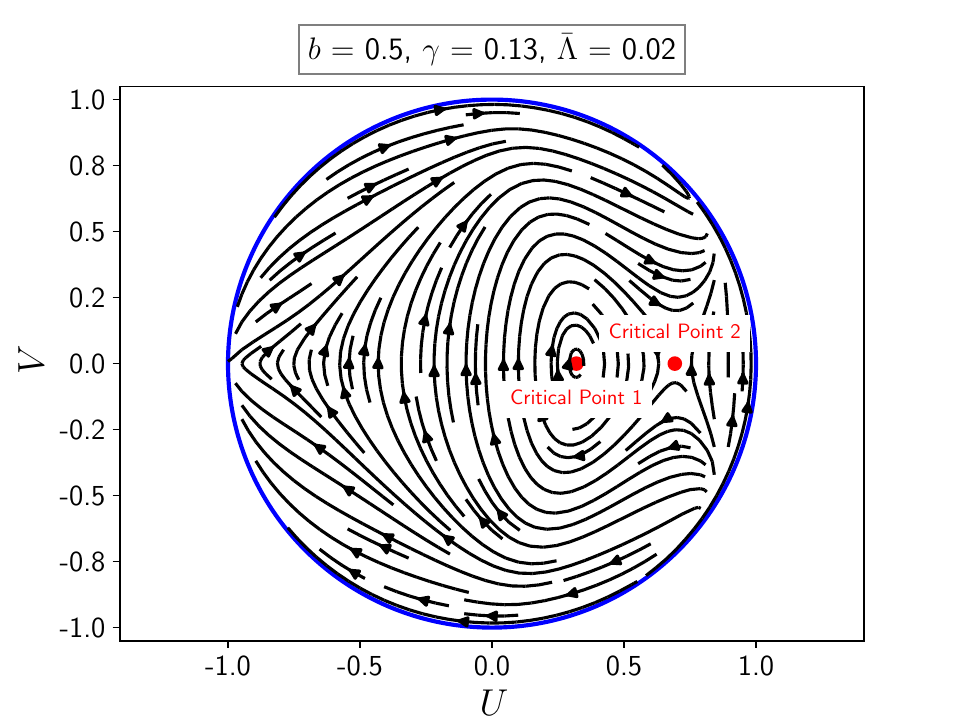}
  \caption{Dynamics of system (\ref{52})-(\ref{53}) for the choice $b=0.5, \gamma=0.13$, and $\overline{\Lambda}=0.02 $.}
  \label{fig:5}
\end{figure}

\noindent Since the system described above is generally unbounded, we introduce the compactification
\begin{eqnarray}
U=\frac{u}{\sqrt{ u^2 +v^2 +1}}, \quad V=\frac{v}{\sqrt{ u^2 +v^2 +1}}. \label{85}\
\end{eqnarray}
The dynamics of the system within the compact plane yields
\begin{eqnarray}
\frac{dU}{d\eta} &=&\frac{UV\sqrt{1-U^2-V^2} e^{\frac{2U}{\sqrt{1-U^2-V^2}}}}{b^2\left[1+e^{\frac{12U}{\sqrt{1-U^2-V^2}}} +e^{\frac{8U}{\sqrt{1-U^2-V^2}}}(3-6\gamma)+e^{\frac{4U}{\sqrt{1-U^2-V^2}}}(3+10\gamma) \right]^2} \nonumber\\
& &\left[\overline{\Lambda}+e^{\frac{24U}{\sqrt{1-U^2-V^2}}}\overline{\Lambda}+e^{\frac{20U}{\sqrt{1-U^2-V^2}}}\left(-1+(6-18\gamma)\overline{\Lambda}\right)+e^{\frac{4U}{\sqrt{1-U^2-V^2}}}\left(3+(6-10\gamma)\overline{\Lambda}\right)\right. \nonumber\\
& & +2e^{\frac{12U}{\sqrt{1-U^2-V^2}}}(3+10\overline{\Lambda}+18\gamma\left(1+3\overline{\Lambda})\right)+e^{\frac{8U}{\sqrt{1-U^2-V^2}}}\left(8+15\overline{\Lambda}+2\gamma(5+24\overline{\Lambda})\right) \nonumber\\
& &\left. +e^{\frac{16U}{\sqrt{1-U^2-V^2}}}\left(15\overline{\Lambda}+\gamma(-6+32\overline{\Lambda})\right)\right] +V(1-U^2-V^2)+V^3, \label{52}\\
\frac{dV}{d\eta} &=&\frac{(V^2-1)\sqrt{1-U^2-V^2} e^{\frac{2U}{\sqrt{1-U^2-V^2}}}}{b^2\left[1+e^{\frac{12U}{\sqrt{1-U^2-V^2}}} +e^{\frac{8U}{\sqrt{1-U^2-V^2}}}(3-6\gamma)+e^{\frac{4U}{\sqrt{1-U^2-V^2}}}(3+10\gamma) \right]^2} \nonumber\\
& &\left[\overline{\Lambda}+e^{\frac{24U}{\sqrt{1-U^2-V^2}}}\overline{\Lambda}+e^{\frac{20U}{\sqrt{1-U^2-V^2}}}\left(-1+(6-18\gamma)\overline{\Lambda}\right)+e^{\frac{4U}{\sqrt{1-U^2-V^2}}}\left(3+(6-10\gamma)\overline{\Lambda}\right)\right. \nonumber\\
& & +2e^{\frac{12U}{\sqrt{1-U^2-V^2}}}(3+10\overline{\Lambda}+18\gamma\left(1+3\overline{\Lambda})\right)+e^{\frac{8U}{\sqrt{1-U^2-V^2}}}\left(8+15\overline{\Lambda}+2\gamma(5+24\overline{\Lambda})\right) \nonumber\\
& &\left. +e^{\frac{16U}{\sqrt{1-U^2-V^2}}}\left(15\overline{\Lambda}+\gamma(-6+32\overline{\Lambda})\right)\right] -UV^2. \label{53}\
\end{eqnarray}
Equations (\ref{52}) and (\ref{53}) provide an overview of the phase space illustrated in Figure \ref{fig:5}. 
We also look for equations around each critical point on the Poincar\'e sphere.

\noindent \textbf{Critical Point 1}
\begin{eqnarray}
U_1 &\approx&\frac{\ln{3}}{\sqrt{16+\ln^2{3}}}+\frac{1024}{9(16+\ln^2{3})^{\frac{3}{2}}}\overline{\Lambda}+\frac{16}{(16+\ln^2{3})^{\frac{3}{2}}}\gamma , \
\end{eqnarray}
\noindent By taking a linear expansion
around critical point 1 and neglecting higher orders terms, we obtain
\noindent \begin{eqnarray}
\frac{dU}{d\eta}=V\left(\frac{16}{16+\ln^2{z_1}} \right), \quad \frac{dV}{d\eta}=-k_1 \left(U-\frac{\ln{z_1}}{\sqrt{16+\ln^2{z_1}}}\right), \label{56}\
\end{eqnarray}
where $z_1 = (x_1)^4$, and
\begin{eqnarray}
\frac{16}{16+\ln^2{z_1}}&\approx& \frac{16}{16+\ln^2{3}}-\frac{2048\ln{3}}{9(16+\ln^2{3})^2}\overline{\Lambda}-\frac{32\ln{3}}{(16+\ln^2{3})^2}\gamma , \nonumber\\
\frac{\ln{z_1}}{\sqrt{16+\ln^2{z_1}}}&\approx&\frac{\ln{3}}{\sqrt{16+\ln^2{3}}}+\frac{1024}{9(16+\ln^2{3})^{\frac{3}{2}}}\overline{\Lambda}+\frac{16}{(16+\ln^2{3})^{\frac{3}{2}}}\gamma , \nonumber\
\end{eqnarray}
and 
\begin{eqnarray}
k_1&\approx&\frac{9\sqrt{3}}{16b^2}+\frac{9\sqrt{3}\ln^2{3}}{256b^2}+\left[-\frac{4\sqrt{3}}{b^2}+\frac{\sqrt{3}\ln{3}}{2b^2}-\frac{\sqrt{3}\ln^2{3}}{4b^2}\right]\overline{\Lambda}+\left[\frac{81\sqrt{3}}{64b^2}+\frac{9\sqrt{3}\ln{3}}{128b^2}+\frac{81\sqrt{3}\ln^2{3}}{1024b^2} \right]\gamma . \nonumber\
\end{eqnarray}
The value of \( k_1 \) is positive when \( 0 \leq \gamma \leq 2.15 \) and \( \overline{\Lambda}_{\text{max}} \geq \overline{\Lambda} > 0 \). Given the initial values \( U(0) = \delta_U + \frac{\ln{z_1}}{\sqrt{\ln^2{z_1} + 16}} \) and \( V(0) = \delta_V \), the solution to equation (\ref{56}) when \( k_1 > 0 \) is given by
\begin{eqnarray}
U(\eta)&=&\frac{\ln{z_1}}{\sqrt{\ln^2{z_1}+16}}+\delta_U \cos {\left(\frac{4\eta\sqrt{k_1}}{\sqrt{\ln^2{z_1}+16}} \right)}+\frac{4\delta_V \sin{\left(\frac{4\eta\sqrt{k_1}}{\sqrt{\ln^2{z_1}+16}} \right)}}{\sqrt{k_1}\sqrt{\ln^2{z_1}+16}}, \label{Ucp1}\\
V(\eta)&=&-\frac{1}{4}\delta_U \sqrt{k_1}\sqrt{\ln^2{z_1}+16}\sin{\left(\frac{4\eta \sqrt{k_1}}{\sqrt{\ln^2{z_1}+16}}\right)}+\delta_V \cos{\left(\frac{4\eta \sqrt{k_1}}{\sqrt{\ln^2{z_1}+16}}\right)}. \label{Vcp1}\nonumber\\
\end{eqnarray}
Solution \eqref{Ucp1}-\eqref{Vcp1} approximates the exact solutions of the full system around critical point 1. The orbits near critical point 1 can be approximated by the ellipses:
\begin{eqnarray}
\left(U-\frac{\ln{z_1}}{\sqrt{\ln^2{z_1}+16}} \right)^2+\frac{16V^2}{k_1(\ln^2{z_1}+16)}&=&\delta_U^2+\frac{16\delta_V^2}{k_1(\ln^2{z_1}+16)}.\
\end{eqnarray}

\noindent \textbf{Critical Point 2}
\begin{eqnarray}
U_2 &\approx&-\frac{96}{(16+\ln^2{\overline{\Lambda}})^{\frac{3}{2}}}\overline{\Lambda}-\frac{\ln{\overline{\Lambda}}}{\sqrt{16+\ln^2{\overline{\Lambda}}}}, 
\end{eqnarray}
\noindent In the same way as for \eqref{56}, the system near the critical point 2 is
\begin{eqnarray}
\frac{dU}{d\eta}=V\left(\frac{16}{16+\ln^2{z_2}} \right), \quad \frac{dV}{d\eta}=-k_2 \left(U-\frac{\ln{z_2}}{\sqrt{16+\ln^2{z_2}}}\right), \label{445}\
\end{eqnarray}
where $z_2 = (x_2)^4$, and
\begin{eqnarray}
\frac{16}{16+\ln^2{z_2}}&\approx& \frac{16}{16+\ln^2{\frac{1}{\overline{\Lambda}}}}+\frac{192\ln{\frac{1}{\overline{\Lambda}}}}{\left(16+\ln^2{\frac{1}{\overline{\Lambda}}}\right)^2}\overline{\Lambda}, \nonumber\\
\frac{\ln{z_2}}{\sqrt{\ln^2{z_2}+16}}&\approx&\frac{\ln{\frac{1}{\overline{\Lambda}}}}{\sqrt{16+\ln^2{\frac{1}{\overline{\Lambda}}}}}-\frac{96}{(16+\ln^2{\frac{1}{\overline{\Lambda}}})^{\frac{3}{2}}}\overline{\Lambda}, \nonumber\
\end{eqnarray}
and
\begin{eqnarray}
k_2=\sqrt{\frac{16}{16+\ln^2 {z_2}}}\left[\left.\frac{\partial^2 W(U)}{\partial U^2}\right|_{U_2,0}\right]. \nonumber\
\end{eqnarray}
The value of \( k_2 \) is negative when \( 0 \leq \gamma \leq 2.15 \) and \( \overline{\Lambda}_{\text{max}} \geq \overline{\Lambda} > 0 \). Given the initial values \( U(0) = \delta_U + \frac{\ln{z_2}}{\sqrt{\ln^2{z_2} + 16}} \) and \( V(0) = \delta_V \), the solution to the above equation for \( k_2 < 0 \) is given by
\begin{eqnarray}
U(\eta)&=&\frac{\ln{z_2}}{\sqrt{\ln^2{z_2}+16}}+\delta_U \cosh {\left(\frac{4\eta\sqrt{-k_2}}{\sqrt{\ln^2{z_2}+16}} \right)}+\frac{4\delta_V \sinh{\left(\frac{4\eta\sqrt{-k_2}}{\sqrt{\ln^2{z_2}+16}} \right)}}{\sqrt{-k_2}\sqrt{\ln^2{z_2}+16}} ,\\
V(\eta)&=&\frac{1}{4}\delta_U \sqrt{-k_2}\sqrt{\ln^2{z_2}+16}\sinh{\left(\frac{4\eta \sqrt{-k_2}}{\sqrt{\ln^2{z_2}+16}}\right)}+\delta_V \cosh{\left(\frac{4\eta \sqrt{-k_2}}{\sqrt{\ln^2{z_2}+16}}\right)}. \nonumber\\
\end{eqnarray}

\noindent The orbits near critical point 2 with \( k_2 < 0 \) can be approximated by hyperbole:
\begin{eqnarray}
\left(U-\frac{\ln{z_2}}{\sqrt{\ln^2{z_2}+16}} \right)^2-\frac{16V^2}{|k_2|(\ln^2{z_2}+16)}&=&\delta_U^2-\frac{16\delta_V^2}{|k_2|(\ln^2{z_2}+16)}.\
\end{eqnarray}

\section{Observational Constraints}
\label{sec:observational constraints}

To determine the parameters $b$, $\gamma$, and $\overline{\Lambda}$ in our cosmological model, we use 44 data points from Cosmic Chronometer observations within the red shift range $0.07 \leq z \leq 2.36$ \cite{bhardwaj2022corrected}. To fit the model, we employ a Bayesian approach and Markov Chain Monte Carlo (MCMC) simulations using the \texttt{emcee} Python library, with our code available at \cite{leonardokap}. This method allows us to rigorously estimate the parameter values by maximizing the likelihood function.

Our MCMC study is carried out with 30,000 walkers and 1,000 steps, using logarithmic priors on the parameters: $b \in [10^{-50}, 10^{-49}]$, $\gamma \in [10^{-3}, 10^{-1}]$, and $\overline{\Lambda} \in [10^{-99}, 10^{-98}]$. From this analysis, we obtain the best-fit values: 
\begin{equation}
b = (2.67 \pm 5.2) \times 10^{-50}, \quad \gamma = (5.0 \pm 3.3) \times 10^{-2}, \quad \text{and} \quad \overline{\Lambda} = (6.13 \pm 2.9) \times 10^{-99}.
\end{equation}
\begin{figure}[ht]
    \centering
    \includegraphics[width=0.8\linewidth]{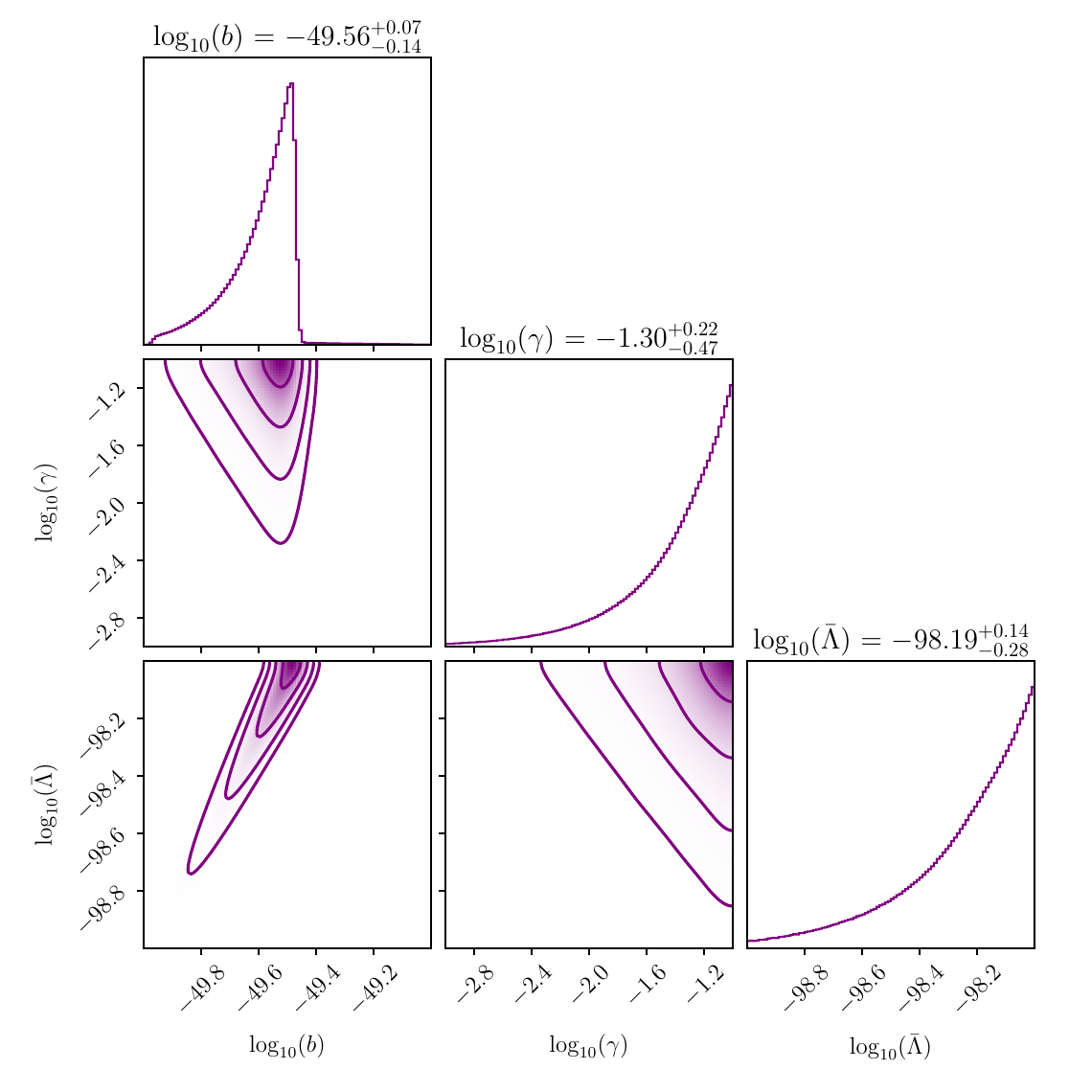}
    \caption{The 1D marginalized posterior distributions and the 2D 68\% and 95\% confidence levels for the $b$, $\gamma$, and $\overline{\Lambda}$ parameters of the cosmological model obtained from OHD data in log scale.}
    \label{fig:corner_plot}
\end{figure}
We use the value of $H_0 = 67.4 \pm 0.5$ km/s/Mpc from the Planck measurements \cite{aghanim2020planck}. Using the best-fit values of the model parameters, we derive the following cosmological parameters: the nonlinear electrodynamics parameter $\beta = 1.56 \times 10^{-116} \, \text{m} \cdot \text{s}^2/\text{GeV}$, the nonminimal coupling parameter $\xi = 2.39 \times 10^{-65} \, \text{s}^2$, the current magnetic field $B_0 = 3.03 \times 10^8 \, \text{GeV}^{1/2}/\text{m}^{1/2}/\text{s}$, and the cosmological constant $\Lambda = 1.28 \times 10^{-35} \, \text{s}^{-2}$. Remarkably, this value of $\Lambda$ aligns well with earlier measurements by the High-Z Supernova Team and the Supernova Cosmology Project, which provided direct evidence of a non-zero cosmological constant \cite{riess1998observational}, with $\Lambda \approx 10^{-122}$ in Planck units. Additionally, it is close to the predicted value from \cite{Moshe}, which gives $\Lambda = 2.03 \times 10^{-35} \, \text{s}^{-2}$. Table \ref{table:comparison} shows the comparison between the value of the cosmological parameters from our model and the references.

\begin{figure}[ht]
    \centering
    \includegraphics[width=0.6\linewidth]{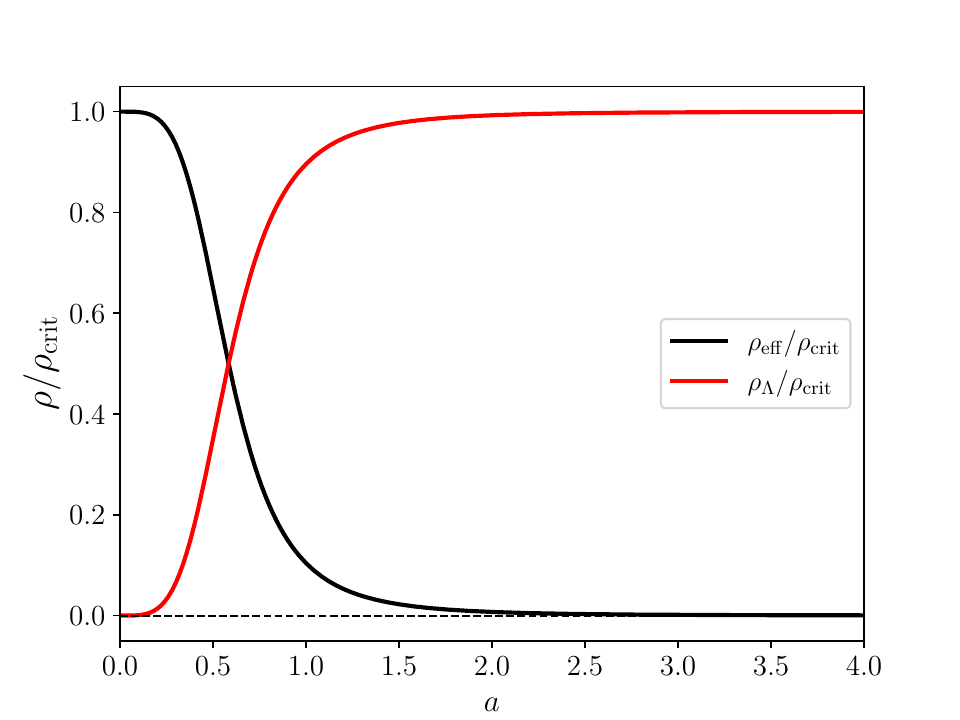}
    \caption{The evolution of the density of the dark energy (DE) component versus scale factor $a$.}
    \label{fig:Energy_density}
\end{figure}

Furthermore, we compute the density parameters for the cosmological constant ($\Omega_\Lambda$) and for nonlinear electrodynamics ($\Omega_{\text{eff}}$) as shown in Fig. \ref{fig:Energy_density}. These are derived from the critical density $\rho_{\text{crit}} = 3H_0^2/\kappa^2$. For $a = 1$, we find $\Omega_{\Lambda,0} = 0.89$ and $\Omega_{\text{eff}, 0} = 0.10$. This suggests that the effective energy density from nonlinear electrodynamics decreases over time as the universe expands, while the dark energy density remains constant. Consequently, the ratio of the effective energy density to the critical density diminishes.

Figure \ref{fig:Energy_density} highlights how the density parameters from the cosmological constant and nonlinear electrodynamics evolve. In the early universe, nonlinear electrodynamics dominates, driving inflation. However, as the universe expands, its influence diminishes, allowing the cosmological constant to become the dominant force. Eventually, as $a \rightarrow \infty$, the cosmological constant becomes the primary contributor to the density parameter. This interplay between energy sources throughout cosmic history illustrates the shift from an inflationary phase to the current acceleration driven by the cosmological constant.

Understanding the dynamics of cosmic expansion also requires knowledge of the Hubble parameter. The Hubble parameter $H(t)$ indicates the rate of the universe's expansion. By using the relation $a = 1/(1+z)$, we obtain the expression for $H(z)$ in terms of redshift $z$:
\begin{equation}
H(z) = H_0 \sqrt{\frac{(1+b^2)^3 - 2\gamma b^2 (3-5b^2)}{b^2(1+b^2) + \overline{\Lambda}(1+b^2)^3}} \sqrt{\frac{b^2(1+z)^4[1+b^2(1+z)^4] + \overline{\Lambda}[1+b^2(1+z)^4]^3}{[1+b^2(1+z)^4]^3 - 2\gamma b^2(1+z)^4[3-5b^2(1+z)^4]}}
\end{equation}
The behavior of $H(z)$ provides valuable insights into the dynamics of the universe's expansion.

\begin{figure}[ht]
    \centering
    \includegraphics[width=.75\linewidth]{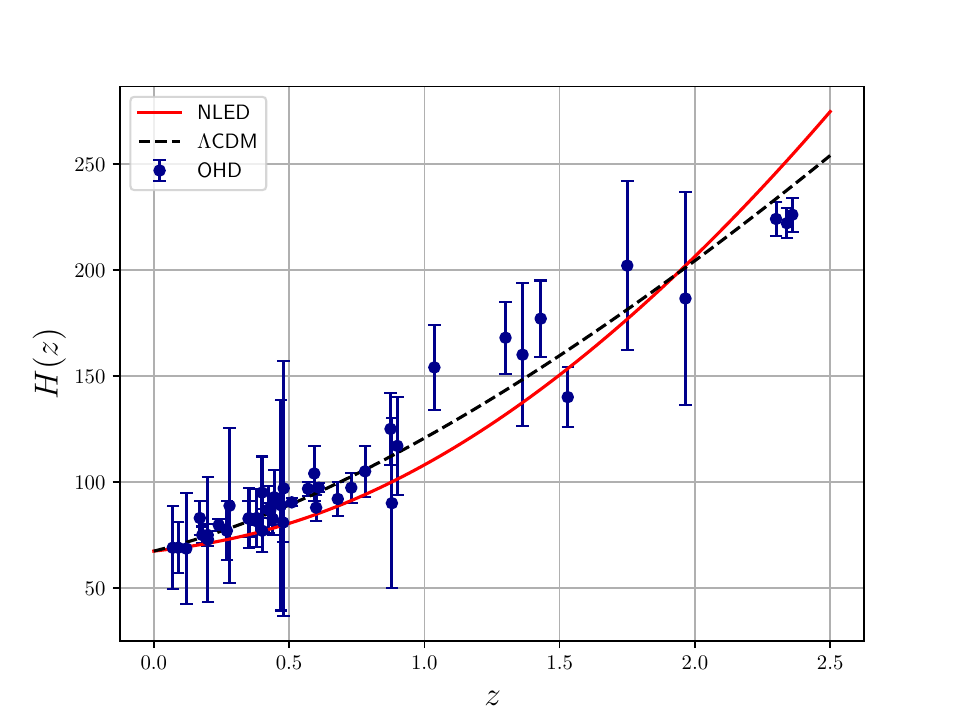}
    \caption{The Hubble parameter as a function of redshift $z$. The red solid curve corresponds to the best-fit values of the model parameters, with blue dots representing observational $H(z)$ data points, and the $\Lambda$CDM model is shown by the dashed line.}
    \label{fig:Hz_vs_z}
\end{figure}

Figure \ref{fig:Hz_vs_z} compares the evolution of the Hubble parameter in our model with the $\Lambda$CDM model. We observe that at redshifts greater than $z \sim 2$, the model predicts a slightly lower Hubble parameter compared to the $\Lambda$CDM model, indicating less deceleration. This close alignment with observational data suggests that our model provides a compelling alternative for explaining the universe's accelerated expansion at later times. However, more observational evidence is required to determine which model best fits the universe's behavior.

As highlighted by cosmological observations, the accelerated expansion of the universe is a relatively recent phenomenon. Before this, the universe underwent a decelerating phase during its early nonlinear electrodynamics-dominated era, which facilitated structure formation. Thus, any cosmological model must account for both decelerated and accelerated phases to fully describe the universe's evolution, a task that our model achieves by integrating nonlinear electrodynamics and a cosmological constant.

The deceleration parameter, $q$, is a critical quantity for understanding the dynamics of the universe’s expansion. It is defined as $q = - \frac{\ddot{a} a}{\dot{a}^2}$, where a negative $q$ indicates acceleration ($\ddot{a}>0$), while a positive $q$ corresponds to deceleration ($\ddot{a}<0$). By utilizing the first and second Friedmann equations \eqref{f1}-\eqref{f2}, along with the dimensionless parameters $x = a \left( 2/\beta B_0^2 \right)^{1/4}$, $\gamma = \xi\kappa^2 / \beta$, and $\overline{\Lambda} = \beta \Lambda / \kappa^2$, we can derive an expression for the deceleration parameter as:
\begin{eqnarray}
q &=& \frac{x^4}{x^4 (x^4 + 1) + \overline{\Lambda}(x^4 + 1)^3} 
\left[ \frac{(x^4 - 3)(x^4 + 1)^3 - 2\gamma x^4 (5 + 18 x^4 - 3 x^8)}{(x^4 + 1)^3 - 2\gamma x^4 (3 x^4 - 5)} \right] \nonumber \\
& &  - \frac{(x^4 + 1)}{x^4 + \overline{\Lambda}(x^4 + 1)^2} 
\left[  \frac{(x^4 + 1)^4 - 2\gamma x^4 (5 - 34 x^4 + 9 x^8)}{(x^4 + 1)^3 - 2\gamma x^4 (3 x^4 - 5)} \right] \overline{\Lambda}. \label{qx}
\end{eqnarray}

Substituting Eq. \eqref{solusiaproksimasi} into Eq. \eqref{qx}, we obtain the approximate solution:
\begin{eqnarray}
q &\approx& -1 + \frac{2\tau}{\sqrt{\tau^2 + b^2}} - \frac{4 \left( 6\tau^5 + 11b^2\tau^3 + 3b^4\tau \right)}{3b^2 \left( \tau^2 + b^2 \right)^{3/2}} \overline{\Lambda} \nonumber \\
& & + \frac{b^2}{4 (\tau^2 + b^2)^{2}} \left[ 16 (\tau^2 - b^2) + b \pi \sqrt{\tau^2 + b^2} \right. \nonumber \\
& & \left. - 4 \sqrt{\tau^2 + b^2} \left( -2b + \tau + b \arctan{\left[ \frac{\tau + \sqrt{\tau^2 + b^2}}{b} \right]} \right)  \right] \gamma. \label{qaproksimasi}
\end{eqnarray}
In the limit $\tau \xrightarrow[]{} \infty$, Eq. \eqref{qaproksimasi} simplifies to
$q \xrightarrow[]{} -\frac{8\tau^2}{b^2} \overline{\Lambda}$, while for $\tau \xrightarrow[]{} 0$, we find $q \xrightarrow[]{} -1 - 2\gamma$. These results demonstrate that our model effectively describes both the early-time inflationary phase, driven by the nonlinear electrodynamics field, and the late-time accelerated expansion, governed by the cosmological constant.

We also explore the duration of inflation and deceleration by employing Eqs. \eqref{Ti}, \eqref{Td} and introducing the time rescaling $\tau = \kappa B_0 t/\sqrt{6}$. Our calculations indicate that the inflationary period lasted approximately $t_i \approx 2. \times 10^{-32}$ seconds—a brief but significant epoch. Following this, the universe transitioned to a decelerating expansion phase, which endured for around $t_d \approx 2.02 \times 10^{17}$ seconds. This deceleration period aligns with the findings of \cite{turner2002type}, after which the universe re-entered an accelerated expansion phase.

At present, our model predicts a deceleration parameter of $q_0 = -0.79$, consistent with recent observations. A study by \cite{camarena2020local} using type Ia supernovae (SN Ia) data in the redshift range $0.023 \leq z \leq 0.15$ reports $q_0 = -1.08 \pm 0.29$, which is in close agreement with our findings. Another work by \cite{giostri2012cosmic}
, combining SN Ia data with baryonic acoustic oscillations (BAO) and cosmic microwave background (CMB) measurements, estimated $q_0 = -0.53^{+0.17}_{-0.13}$. These small differences reflect the various methodologies and datasets used to constrain the deceleration parameter, further highlighting the need for continued observational efforts to refine these estimates.

\begin{table}[!ht]
    \centering
    \renewcommand{\arraystretch}{1.5} 
    \begin{tabular}{|c|c|cc|}
    \hline
    Parameter & Model & \multicolumn{2}{c|}{Reference} \\ \hline
    $\Lambda$ & $1.28\times 10^{-35}$ s$^{-2}$  & \multicolumn{1}{c|}{$2.03 \times 10^{-35}$ s$^{-2}$} & \cite{Moshe}  \\ \hline
    $\Omega_{\Lambda}$ & $0.89$  & \multicolumn{1}{c|}{$0.6847 \pm 0.0073$} & \cite{aghanim2020planck} \\ \hline
    \multirow{2}{*}{$q_0$} & \multirow{2}{*}{$-0.79$} & \multicolumn{1}{c|}{$-1.08 \pm 0.29$} & \cite{camarena2020local} \\ \cline{3-4} 
     &  & \multicolumn{1}{c|}{$-0.53^{+0.17}_{-0.13}$} & \cite{giostri2012cosmic} \\ \hline
    $t_i$ & $1.15\times 10^{-32}$ s & \multicolumn{1}{c|}{$10^{-32}$ s} & \cite{kruglov2024universe} \\ \hline
    \multirow{2}{*}{$t_d$} & \multirow{2}{*}{$2.02 \times 10^{17}$ s} & \multicolumn{1}{c|}{$2.02 \times 10^{17}$ s} & \cite{turner2002type} \\ \cline{3-4} 
     &  & \multicolumn{1}{c|}{$\sim 2 \times 10^{17}$ s} & \cite{akarsu2017inflation} \\ \hline
    \end{tabular}
    \caption{The value comparison of the cosmological parameters between our model and references.}
    \label{table:comparison}
\end{table}

\section{Conclusions}
\label{sec:conclusion}
In this work, we have studied a nonminimally coupled gravity model with nonlinear electrodynamics to explain the inflation of the universe in its early stages and accelerated expansion in its late time of evolution. An inflationary universe is necessary for solving several puzzles in standard Big Bang cosmology, such as the monopole, horizon, and flatness problems. Using a nonlinear electrodynamics model, we have demonstrated that the early universe was filled with a remarkably stochastic magnetic field, which played a crucial role in driving the universe to inflate. The presence of strong magnetic fields at the beginning of the inflationary period can impact the behavior of inflation and generate primordial density fluctuations, which are the seeds for the large-scale structure of the universe we observed today. Furthermore, we found no singularities in the energy density and pressure within our nonminimal coupling model as $a(t) \rightarrow 0$ and $a(t) \rightarrow \infty$. The absence of singularities is an interesting feature of cosmological models with nonlinear electrodynamics.

When combining nonminimal coupling gravity with a nonlinear electrodynamics model alone, we found it is not possible to generate a solution for a universe that experiences both early-time inflation and late-time accelerated expansion. Therefore, we added a cosmological constant, $\Lambda$, to take into account  the late-time accelerated expansion. By defining the dimensionless parameters $ \gamma\equiv \xi \kappa^2/\beta$ and $\overline{\Lambda}\equiv\beta \Lambda/\kappa^2$, where $\xi$ and $\beta$ are the nonminimal coupling parameter and the nonlinear electrodynamics parameter, respectively, we found that the effective potential $W(x)$ is valid within the range $\gamma \leq 2.15$. The universe accelerates as the potential function decreases with respect to $x=a/\sqrt{b}$. Our proposed model can be used as a toy model to explore various scenarios of cosmic acceleration by varying the parameters $\gamma$ and  $\overline{\Lambda}$ at the initial time. This allows us to analyze the different phases of the universe's evolution, including the duration of inflation and periods of deceleration. A notable feature of this model is the existence of critical points that indicate phases of inflation, deceleration, and late-time acceleration of the universe.
  
We have tested our model by performing Bayesian analysis and Monte Carlo (MCMC) simulations to a collection of 44 data points obtained from the Cosmic Chronometer observations in the redshift range $0.07 \leq z \leq 2.36$. We obtained the nonlinear electrodynamics parameter \(\beta = 4.26 \times 10^{-117} \, \text{m} \cdot \text{s}^2/\text{GeV}\), the nonminimal coupling parameter \(\xi = 6.80 \times 10^{-67} \, \text{s}^2\), the current value of the magnetic field \(B_0 = 3.03 \times 10^8 \, \text{GeV}^{1/2}/\text{m}^{1/2}/\text{s}\), and the cosmological constant \(\Lambda = 1.28 \times 10^{-35} \, \text{s}^{-2}\). Using these parameters, we found that the duration of inflation is approximately \(t_i = 1.15 \times 10^{-32}\) s, the duration of deceleration is approximately \( t_d = 2.02 \times 10^{17} \), and the value of deceleration parameter today is $q_0=-0.79$.

In conclusion, the nonlinear electrodynamics models have been widely studied and successfully explained the early inflationary universe, but the models alone cannot account for late-time cosmic acceleration. Conversely, the $\Lambda$CDM model can only explain the late-time acceleration and fails to produce two distinct periods of acceleration in early-time and late-time. By coupling nonlinear electrodynamics and gravity nonminimally and considering the cosmological constant, we successfully developed a model that explains both the early inflationary period and the late-time accelerated expansion of the universe. From the deceleration parameter $q$, we found that $q\xrightarrow[]{}-1-2\gamma$ as $\tau \xrightarrow[]{}0$ and $q \xrightarrow[]{} -\frac{8\tau^2}{b^2}\overline{\Lambda}$ as $\tau \xrightarrow[]{}\infty$. These results indicate that our model provides solutions for a universe experiencing early-time inflation driven by the dominance of the nonlinear electrodynamics field and late-time accelerated expansion driven by the dominance of the  cosmological constant.

\section*{Acknowledgments}
FTA and BEG acknowledge ITB Research Grant 2024 for financial support.


\printbibliography
\end{document}